\documentclass[11pt]{article}
\usepackage{moriond,epsfig}

\def\Journal#1#2#3#4{{#1} {\bf #2}, #3 (#4)}

\def\NIMA{{\em Nucl. Instrum. Methods} A}
\def\NPB{{\em Nucl. Phys.} B}
\def\PLB{{\em Phys. Lett.}  B}
\def\PRL{\em Phys. Rev. Lett.}
\def\PRD{{\em Phys. Rev.} D}

\def\be{\begin{equation}}
\def\ee{\end{equation}}
\def\bea{\begin{eqnarray}}
\def\eea{\end{eqnarray}}

\begin{document}
\title{Charm and $\tau$ Decays: Review of BaBar and Belle Results}

\author{Jens S\"oren Lange\\ Email soeren.lange@exp2.physik.uni-giessen.de }

\address{Justus-Liebig-Universit\"at Giessen, II.\ Physikalisches Institut, 
Heinrich-Buff-Ring 16, 35392 Giessen}

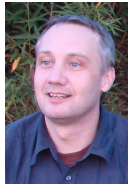
\maketitle\abstracts{A review of recent results concerning decays of charmed mesons, charmed baryons
and $\tau$ leptons for the Belle and BaBar Experiments is given. The review was presented at the
43$^{rd}$ Rencontres de Moriond (QCD and High Energy Interactions), La Thuile, March 8-15, 2008.}

\vspace*{-0.5cm}

\noindent
The $B$ meson factories can in fact also be considered as  
charm and $\tau$ factories, as 
{\it (a)} $\simeq$99\% of all $B$ mesons decay to charm final states, and
{\it (b)} the cross sections of charm and $\tau$ production in the continuum, 
i.e.\ $e^+$$e^-$$\rightarrow$$q$$\overline{q}$ ($q$=$u$,$d$,$s$,$c$)
and $e^+$$e^-$$\rightarrow$$\tau^+$$\tau^-$, 
are as high as the cross section for $B$ meson production ($\simeq$1~nb)
in the $Y$(4S) decay. Details of the BaBar and Belle detectors can be found 
elsewhere \cite{babar_nim}$^{, }$\cite{belle_nim}. At the time of this review, the integrated 
luminosities are 512.2~fb$^{-1}$ for BaBar and 773.7~fb$^{-1}$ for Belle.

\section{Charm Decays}
\label{}

\subsection{$D_{sJ}$ states}

\noindent
Earlier two new $D_{sJ}$ states have been observed: 
the $D_{s0}$(2317) by BaBar \cite{babar_dsj2317} 
in the decay\footnote{Throughout this paper 
charge conjugations are implied if not noted otherwise.} 
$D_{s0}^{+}$(2317)$\rightarrow$$D_s^+$$\pi^0$, 
and the $D_{s1}$(2460), by CLEO \cite{cleo_dsj2460}
in the decay $D_{s1}^{+}$(2460)$\rightarrow$$D_s^{*+}$$\pi^0$.
In both cases, due to the $D_s$ in the final state,
the most probable assignment is a $($$c$$\overline{s}$$)$ 
state with $L$=1. However, the measured masses were found
$\simeq$100~MeV too low compared to early quark models \cite{isgur_weise},
and inspired more theoretical and experimental work.
Recent analyses have investigated 
decays of higher-mass $D_{sJ}$ states
not only into $D_s$$\pi$, but also into $D^{(*)}$$K$,  
i.e.\ corresponding to a Feynman diagram with an 
internally (rather than externally) created $u$$\overline{u}$ pair.

\subsection{$D_{sJ}$ decays into $D^*$$K$}

This category of decays (assuming $L$=0) favors the production of 
vector or axial-vector $D_{sJ}$ states
due to the $D^*$ vector meson in the final state.
Recently BaBar reported the first observation
of $D_{s1}$(2536) in $B$ decays.
A data set of 347 fb$^{-1}$ was used to investigate 
the decays $D_{s1}$(2536)$\rightarrow$$D^{*+}$$K_s^0$ and 
$D_{s1}$(2536)$\rightarrow$$D^{*0}$$K^+$.
An analysis of the $D_{s1}$(2536) helicity was performed
in order to confirm the historically assigned PDG \cite{PDG} quantum numbers
of $J^P$=1$^+$, which is the natural expectation in the heavy charm-quark limit
for the lower member of an $L$=1, $j_q$=3/2 doublet.
$J^P$=1$^+$ in a pure S-wave as well as 
$J^P$=2$^+$ and $J^P$=2$^-$
were disfavoured, whereas 
fits for $J^P$=1$^-$ in a pure P-wave and 
$J^P$=1$^+$ with S/D-wave admixture both 
describe the observed angular distribution well.
On the other hand, Belle reported an analysis of $D_{s1}$(2536)$\rightarrow$$D^{*+}$$K_s^0$,
but using continuum production $e^+$$e^-$$\rightarrow$$D_{s1}$(2536)$X$.
A partial wave analysis on a data set of 462 fb$^{-1}$ was performed,
in order to study the mixing of the
($j_q$=1/2) $D_{s1}$(2460) and the ($j_q$=3/2) $D_{s1}$(2536) states.
Heavy Quark Effective Theory (HQET) predicts, 
that ($j_q$=3/2)$\rightarrow$$D^*$$K$ 
should be a pure D-wave decay, and
($j_q$=1/2)$\rightarrow$$D^*$$K$ 
should be a pure S-wave decay.
If HQET is not exact, mixing between the S and D-waves would be possible
by LS interaction:\\
$|$$D_{s1}$(2460)$>$=cos$\theta$$|$$j_q$=1/2$>$+sin$\theta$$|$$j_q$=3/2$>$
and 
$|$$D_{s1}$(2536)$>$=$-$sin$\theta$$|$$j_q$=1/2$>$+cos$\theta$$|$$j_q$=3/2$>$.\\
The fit result yields 
an admixture of $D$/$S$= (0.63$\pm$0.07(stat.))$\cdot$ exp($\pm$$i$(0.77$\pm$0.03(stat.))).
The result indicates that the S-wave dominates in the decay
$D_{s1}$(2536)$\rightarrow$$D^*$$K$ with a fraction of
(72$\pm$3(stat.)$\pm$1(syst.))\%, which contradicts HQET.
The reason might be given by the fact, that 
HQET assumes the $c$ quark to be infinitely heavy.
However, it should be noted that the interpretation is not trivial,
as the D-wave might also be suppressed by the centrifugal barrier.
In any case, there is strong indication that the $D_{s1}$(2460) and the $D_{s1}$(2560) 
in fact are mixing, which might be considered surprising for two very narrow states
($\Gamma$$\leq$1~MeV) with a mass difference of $\Delta$$m$$\simeq$76~MeV.
As a by-product of this analysis, 
Belle reported the first observation of a three-body decay mode of the $D_{sJ}$(2536),
in the decay $D_{s1}$(2536)$\rightarrow$$D^{*0}$$K^+$$\pi^-$.

\subsection{$D_{sJ}$ decays into $D$$K$}

This category of decays (assuming $L$=0) favors the production of 
scalar $D_{sJ}$ states
due to the two pseudoscalar mesons in the final state.
Recently Belle reported the observation of a new $D_{sJ}$ meson,
tentatively called $D_{sJ}$(2700), 
in $B^+$$\rightarrow$$\overline{D}^0$$D^0$$K^+$ decays
with a data set of 414 fb$^{-1}$.
The $D_{sJ}$(2700) was found to be the dominating resonance 
in this $B$ decay. The mass was determined as 
$m$=2708$\pm$9(stat.)$^{+11}_{-10}$(syst.)~MeV,
the width as $\Gamma$=108$\pm$23(stat.)$^{+36}_{-31}$(syst.)~MeV.
The signal yield was reported as 182$\pm$30 events 
with a statistical significance of 8.4$\sigma$.
In order to determine the quantum numbers,
a helicity analysis was performed.
A fit to the distribution of the helicity angle 
between the $D^0$ and the $D_{sJ}$
prefers an assignment of $J$=1.
In this case, a $J$=1$\rightarrow$$0^-$$0^-$ decay
would imply $L$=1, and thus a negative parity assignment.
Two possible interpretations of a $J^P$=$1^-$ $D_{sJ}$(2700) state 
were proposed, i.e.\ {\it (a)} a radial $n$=2 excitation (2$^3S_1$),
predicted \cite{theory_dsj2700_a} by potential models at m$\simeq$2720~GeV,
or {\it (b)} a chiral doublet state $J^P$=$1^-$ as a partner to the $J^P$=$1^+$ $D_{s1}$(2536),
predicted \cite{theory_dsj2700_b} from chiral symmetry considerations at $m$=2721$\pm$10~MeV. 
The new Belle result could be compared to a prior search \cite{babar_dsj2860} 
for higher $D_{sJ}$ states 
in the $e^+$$e^-$ continuum by BaBar with a data set of 240~$fb^{-1}$.
Three structures were observed: 
{\it (a)} the known $D_{s2}$(2573) state, 
{\it (b)} a new $D_{sJ}$(2860) state, and 
{\it (c)} a broad structure peaking around 2.7~GeV, 
which could be identical to the $D_{sJ}$(2700).
As the $D_{sJ}$(2860) is not seen in the Belle data,
this might indicate that probably a higher $J$ should be assigned, 
and thus it is only produced in continuum, but not in $B$ decays.

\subsection{$D_s^+$$\rightarrow$$\mu^+$$\nu_{\mu}$}

\noindent
The purely leptonic decay $D_s^+$$\rightarrow$$\mu^+$$\nu_{\mu}$
is theoretically rather clean, as in the standard model the decay is mediated
by a single virtual W boson. The decay rate is given by

\begin{equation}
\Gamma ( D_s^+ \rightarrow l^+ \nu_l ) = 
\frac{G_F^2}{8\pi}
f_{Ds}^2 m_l^2 m_{Ds} ( 1 - \frac{m_l^2}{m_{Ds}^2} )^2 | V_{cs} |^2
\end{equation}

\noindent
using the Fermi coupling constant $G_F$, 
the masses of the lepton and of the $D_s$ meson, $m_l$ and $m_{Ds}$, respectively, 
and the CKM matrix element $V_{cs}$.
All effects of the strong interaction are accounted
for by the decay constant $f_{Ds}$. 
The measurement of the branching fraction
allows the determination of $f_{Ds}$ and the comparison to 
theoretical or Lattice QCD calculations.
Recently several measurements of $f_{Ds}$ were performed.
An overview is given in Tab.~1.
BaBar reported \cite{babar_dsmunu} a signal yield of 489$\pm$55(stat.) 
$D_s^+$$\rightarrow$$\mu^+$$\nu_{\mu}$ events in a data sample
of 230~fb$^{-1}$.
Belle reported \cite{belle_dsmunu} a signal yield of 169$\pm$16(stat.)$\pm$8(syst.)
$D_s^+$$\rightarrow$$\mu^+$$\nu_{\mu}$ events in a data sample
of 548 fb$^{-1}$. 
The reconstruction procedures show a few technical differences,
i.e.\ the signal either peaks in {\it (a)} the mass difference 
of $m$($\mu$$\nu$$\gamma$)-$m$($\mu$$\nu$)=143.5~MeV, 
corresponding to the photon energy in the 
$D_s^*$$\rightarrow$$D_s$$\gamma$ transition, or
peaks in {\it (b)} the recoil mass $m$($D$$K$$X$$\gamma$$\mu$)=0, 
where $X$=$n$$\pi$ and $\geq$1$\gamma$ with $n$=1,2,3. 
In addition, while Belle determines an absolute branching fraction, 
the BaBar experiment provides a measurement of the partial width ratio
$\Gamma$($D_s^+$$\rightarrow$$\mu^+$$\nu_{\mu}$/
$\Gamma$($D_s^+$$\rightarrow$$\phi$$\pi^+$).
The recent measurements of $f_{Ds}$ of Belle, BaBar and also CLEO-c \cite{cleo-c_dsmunu} 
are consistent 
within the error estimates. However, theoretical calculations seem to indicate 
a lower value of $f_{Ds}$ \cite{dsmunu_rosner_stone}. As an example, the result of 
a recent Lattice QCD calculation \cite{dsmunu_lattice} is $f_{Ds}$=241$\pm$3~MeV.
The discrepancy
between experiment and theory gave rise to speculation as a possible indication
of new physics. Note that charm quark loops are not included in the Lattice 
QCD calculation.

\begin{table}[t]
\caption{Overview of recent measurements of the branching fraction for
$D_s^+$$\rightarrow$$\mu^+$$\nu_{\mu}$ and $f_{Ds}$.}
\vspace{0.4cm}
\begin{center}
\begin{tabular}{|l|l|l|}
\hline
 & Br($D_s$$\rightarrow$$\mu^+$$\nu_{\mu}$) & $f_{Ds}$ (MeV) \\
\hline
\hline
PDG06 \cite{PDG} & (6.1$\pm$1.9)$\cdot$10$^{-3}$ & \\
\hline
BaBar & (6.74$\pm$0.83$\pm$0.26$\pm$0.66)$\cdot$10$^{-3}$ & 283$\pm$17$\pm$7$\pm
$14\\
\hline
Belle & (6.44$\pm$0.76$\pm$0.57)$\cdot$10$^{-3}$ & 275$\pm$16$\pm$12\\
\hline
CLEO-c \cite{cleo-c_dsmunu} & (5.94$\pm$0.66$\pm$0.31)$\cdot$10$^{-3}$ & 274$\pm$13$\pm$7\\
\hline
\end{tabular}
\end{center}
\end{table}

\subsection{Decays of Charmed Baryons}

\noindent
Recently new studies of the $\Xi_c$ charmed baryon system\footnote{The valence quark contents 
of the $\Xi_c^+$ and the $\Xi_c^0$ are given by $($$u$$s$$c$$)$ and $($$d$$s$$c$$)$, respectively.} 
were performed by both BaBar and Belle.
Formerly, the $\Lambda_c$$K$$\pi$ final state has been used
at the $B$ meson factories 
in searches for double charm baryon ground states. For these $($$c$$c$$q$$)$ states,
this final state indicates {\it weak decays}. 
However, Belle formerly observed \cite{belle_xic_a} two new $\Xi_c^*$ states,
i.e.\ the $\Xi_c(2980)^{+,0}$ and the $\Xi_c(3077)^{+,0}$
in $\Lambda_c^+$$K^-$$\pi^+$ and 
$\Lambda_c^+$$K_s^0$$\pi^-$. 
The doublet nature of these states clearly indicate the $\Xi_c^*$ interpretation.
Contrary to double charm baryons, for $\Xi_c^*$ baryons these final states 
indicates {\it strong decays}.
The discovery of these decays might be considered as a surprise, as
{\it (a)} formerly only $\Xi_c^*$ decays to $\Xi_c$$\gamma$
and $\Xi_c$$\pi$ were observed, and 
{\it (b)} here the charm and the strange quarks are observed in {\it different} 
hadrons. Thus the decay reveals an interesting dynamics.
Recently BaBar \cite{babar_xic} also published an updated $\Xi_c$ analysis 
based upon 384 fb$^{-1}$. 
Besides the confirmation of the $\Xi_c(2980)$ and the $\Xi_c(3077)$,
two new states were observed. In the analysis, the final state of 
$\Lambda_c$$K$$\pi$ (same as above) was required, however using a cut 
on the $\Sigma_c$ invariant mass in the $\Lambda_c$$\pi$ subsystem. 
The new states and observed decays are:
$\Xi_c(3055)^+$$\rightarrow$$\Sigma_c(2455)^{++}$($J^P$=1/2$^+$)$K^-$ 
with a statistical significance of 6.4$\sigma$, and 
$\Xi_c(3122)^+$$\rightarrow$$\Sigma_c(2520)^{++}$($J^P$=3/2$^+$)$K^-$ 
with a statistical significance of 3.6$\sigma$.
The observation of these decays imposes a new question to the understanding 
of the dynamics in the decay,
namely why the $\Xi_c$ states decay into a $\Sigma_c$ with an isospin 1.
Again, the charm and the strange quark are observed in {\it different}
hadrons, i.e.\ ($u$$s$$c$)$\rightarrow$($u$$u$$c$)($\overline{u}$$s$).
Belle recently continued \cite{belle_xic_b} the investigation of the $\Xi_c$(2980)
with a data set of 414 fb$^{-1}$ in order to try
to determine the nature of this state.
The $\Xi_c$(2980) could be the first positive parity excitation
of the $\Xi_c$, or a higher ($n$$\geq$2) radial excitation.
A new decay mode $\Xi_c$(2980)$\rightarrow$$\Xi_c$(2645)$\pi$
was observed. Single pion transitions of such kind can give 
important hints for the assignment of quantum numbers.
For example, the transition
$\Xi_c$(2815)($J^P$=3/2$^-$)$\rightarrow$$\Xi_c$(ground state)$\pi$ 
is forbidden, but the transition
$\Xi_c$(2815)($J^P$=3/2$^-$)$\rightarrow$$\Xi_c$(2645)$\pi$ 
is allowed.
On the other hand, the double pion transition 
$\Xi_c$(2815)($J^P$=3/2$^-$)$\rightarrow$$\Xi_c$(ground state)$\pi$$\pi$
is allowed in any case and thus less indicative for quantum number assignments.
Assuming S-wave, the newly observed decay mode 
is predicted \cite{theory_xic} dominant for the 
$\Xi_{c1}$($J^P$=1/2$^+$) state, and thus might favour
an assignment of a positive parity.


\section{$\tau$ Decays}
\label{}

In the simplest tree diagram, hadronic $\tau$ decays are given 
by a $\tau$$\rightarrow$$\nu$$W$
transition, in which the virtual $W$ boson forms a 
quark anti-quark pair $q$$\overline{q}'$
with $q$ and $q'$ carrying different flavor.
As any additional gluon might be soft,
$\alpha_S$$\simeq$0.35 is quite large, and therefore these 
decays are an interesting tool to study non-perturbative QCD.

\subsection{$\tau$ decays with an $\eta$ in the final state}

Belle recently improved \cite{belle_tau_eta} known branching fractions for 
$\tau$ decays with $\eta$ and two additional pseudoscalar mesons 
by a factor 4-6 with a data set of 485~$fb^{-^1}$.
For the decays 
$\tau^-$$\rightarrow$$K^-$$\pi^0$$\eta$$\nu_{\tau}$
and 
$\tau^-$$\rightarrow$$\pi^-$$\pi^0$$\eta$$\nu_{\tau}$
branching ratios of 
4.7$\pm$1.1(stat.)$\pm$0.4(syst.)$\cdot$10$^{-5}$
and 
4.7$\pm$1.1(stat.)$\pm$0.4(syst.)$\cdot$ 10$^{-3}$
were measured, respectively.
The measurement of the branching fractions is very important 
for understanding of low energy QCD. 
In the case of three pseudoscalar mesons the branching fraction
could be increased by a factor $\geq$10 by the Wess-Zumino-Witten anomaly 
\cite{wess-zumino-witten}.

\subsection{$\tau$ decays with an $\phi$ in the final state}

Recently BaBar published \cite{babar_tau_phi} an analysis of $\tau$ decays
into a $\phi$ and an additional pseudoscalar meson
with a data set of 342 $fb^{-1}$.
The decay $\tau^-$$\rightarrow$$\phi$$K^-$$\nu_{\tau}$
is Cabibbo suppressed by the CKM matrix element $V_{us}$$\simeq$0.2.
The measured branching fraction of 3.39$\pm$0.20(stat.)$\pm$0.28(syst.)$\cdot$10$^{-5}$
is consistent with an earlier Belle result 
\cite{belle_tau_phi}, and thus could be used 
as a reference for the branching fractions
of even more rare processes.
On the one hand, excluding the resonant $\phi$ contribution,
an upper limit for the non-resonant 
$\tau^-$$\rightarrow$$K^+$$K^-$$K^-$$\nu_{\tau}$
of $<$2.5$\cdot$10$^{-6}$ was obtained,
indicating that the $\phi$ largely dominates the decay.
On the other hand, a first measurement of the 
branching fraction of the 
OZI suppressed process $\tau^-$$\rightarrow$$\phi$$\pi^-$$\nu_{\tau}$
was performed with the result of
3.42$\pm$0.55(stat.)$\pm$0.25(syst.)$\cdot$10$^{-5}$.





\section*{References}




\begin{thebibliography}{99}

\bibitem{babar_nim} 
BaBar Collaboration, B. Aubert et al., \Journal{\NIMA}{479}{1}{2002}

\bibitem{belle_nim} 
Belle Collaboration, A.~Abashian et al., \Journal{\NIMA}{479}{117}{2002}

\bibitem{babar_dsj2317} 
BaBar Collaboration, B. Aubert et al., \Journal{\PRL}{90}{242001}{2003}

\bibitem{cleo_dsj2460} 
CLEO Collaboration, D.~Besson et al.,  \Journal{\PRD}{68}{032002}{2003}

\bibitem{isgur_weise} 
S.~Godfrey, N.~Isgur, \Journal{\PRD}{32}{189}{1985} 

\bibitem{babar_dsj2536} 
BaBar Collaboration, B.~Aubert at al., {\em Phys.~Rev.} D-RC {\bf 77}, 011102 (2008)

\bibitem{PDG} 
W.-M. Yao et al., {\em Journal of Physics} G {\bf 33}, 1 (2006)

\bibitem{belle_dsj2536} 
Belle Collaboration, V.~Balagura et al., 
\Journal{\PRD}{77}{032001}{2008}

\bibitem{belle_dsj2700}
Belle Collaboration, J.~Brodzicka et al., \Journal{\PRL}{100}{092001}{2008}

\bibitem{theory_dsj2700_a}
F.~E.~Close, C.~E.~Thomas, O.~Lakhina, E.~S.~Swanson, 
\Journal{\PLB}{647}{2007}{159}

\bibitem{theory_dsj2700_b} 
M.~A.~Nowak, M.~Rho, I.~Zahed, 
{\em Acta Phys.~Polon.} B {\bf 35}, 2377 (2004)

\bibitem{babar_dsj2860} 
BaBar Collaboration, B. Aubert et al., \Journal{\PRL}{97}{222001}{2006}

\bibitem{babar_dsmunu} 
BaBar Collaboration, B. Aubert et al., \Journal{\PRL}{98}{141801}{2007}

\bibitem{belle_dsmunu} 
Belle collaboration, L.~Widhalm et al.,
arXiv:0709.1340$[$hep-ex$]$, subm.\ to {\em Phys.\ Rev.\ Lett.\ }

\bibitem{cleo-c_dsmunu}
CLEO-c Collaboration, T.~K.~Pedlar et al.,
\Journal{\PRD}{76}{072002}{2007}

\bibitem{dsmunu_rosner_stone} For a recent review see J.~L.~Rosner,
S.~Stone, arXiv:0802.1043$[$hep-ex$]$

\bibitem{dsmunu_lattice} 
E.~Follana, C.~T.~H.~Davies, G.~P.~Lepage, J.~Shigemitsu\\ 
(HPQCD and UKQCD Collaborations), \Journal{\PRL}{100}{062002}{2008}

\bibitem{belle_xic_a} 
Belle Collaboration, R.~Chistov et al.
\Journal{\PRL}{97}{162001}{2006}

\bibitem{babar_xic} 
BaBar Collaboration, B. Aubert et al., 
\Journal{\PRD}{77}{012002}{2008}

\bibitem{belle_xic_b} 
Belle Collaboration, T.~Lesiak et al
arXiv:0802.3968$[$hep-ex$]$, subm.\ to {\em Phys.\ Lett.\ B}

\bibitem{theory_xic}
H.-Y.~Cheng, C.-K.~Chua, \Journal{\PRD}{75}{014006}{2007} 

\bibitem{belle_tau_eta} 
The Belle Coll., K.~Abe et al., arXiv:0708.0733$[$hep-ex$]$

\bibitem{wess-zumino-witten}
J.~Wess, B.~Zumino, \Journal{\PLB}{37}{95}{1971};\\
E.~Witten, \Journal{\NPB}{223}{422}{1983}

\bibitem{babar_tau_phi}
BaBar Collaboration, B.~Aubert et al., 
\Journal{\PRL}{100}{011801}{2008}

\bibitem{belle_tau_phi}
K.~Inami et al., \Journal{\PLB}{643}{5}{2006}

\end{thebibliography}
\end{document}